\documentclass{article}

\usepackage[T1]{fontenc}
\usepackage[latin1]{inputenc}
\usepackage{epsfig}
\usepackage{amsmath,amssymb}

\def\ctf{C_{\rm TF}}
\def\cxc{C_{\rm xc}}
\def\RR{\mathbb{R}}
\def\ZZ{\mathbb{Z}}

\newcommand \dps {\displaystyle}

\begin{document}
\numberwithin{equation}{section}
\newtheorem{theoreme}{Theorem}[section]
\newtheorem{proposition}[theoreme]{Proposition}
\newtheorem{remarque}[theoreme]{Remark}
\newtheorem{lemme}[theoreme]{Lemma}
\newtheorem{corollaire}[theoreme]{Corollary}
\newtheorem{definition}[theoreme]{Definition}

\title{Nonlinear instability of density-independent \\
orbital-free kinetic energy functionals}
\author{X. Blanc$^1$, E. Cancès$^2$
\\
\\
{\footnotesize $^1$ Laboratoire Jacques-Louis Lions,}\\
{\footnotesize Universit\'e Paris 6, Bo\^{\i}te courrier 187}\\
{\footnotesize 75252 Paris cedex 05, FRANCE}\\
\\
{\footnotesize $^2$ CERMICS, \'Ecole Nationale des Ponts et
Chauss\'ees and INRIA,}\\
{\footnotesize 6 \& 8, avenue Blaise Pascal, Cit\'e Descartes,}\\
{\footnotesize Champs~sur~Marne, 77455 Marne-La-Vall\'ee Cedex 2, FRANCE}\\
}

\maketitle
\begin{abstract}
We study in this article the mathematical properties of a class of
orbital-free kinetic energy functionals. We prove that these models are
linearly stable but nonlinearly unstable, in the sense that the
corresponding kinetic energy functionals are not bounded from below. As
a matter of illustration, we provide an example of an electronic density
of simple shape the kinetic energy of which is negative.
\end{abstract}

\section{Introduction}

\noindent
Kohn-Sham models have brought a considerable breakthrough in
atomic-scale simulation of materials in condensed phase. However, the
use of the Kohn-Sham kinetic energy is problematic when the number of
electrons by unit cell exceeds a few hundreds (for computational means
available to date). Some authors therefore proposed to approximate
the Kohn-Sham kinetic energy functional in order to get rid of both the
orbital and $k$-point dependencies. Their approach consists in  
improving the Thomas-Fermi model, for which the kinetic energy
functional reads 
\begin{equation}
  \label{eq:TF}
  T_{\rm TF}[\rho] = \ctf \int_Q \rho^{5/3} 
\end{equation}
where $Q$ is the simulation unit cell, $\rho$ a given density, and
$\ctf = \dps \frac{3}{10} \left( 3\pi^2 \right)^{2/3}$ the 
Thomas-Fermi constant, by adding some correction terms. The functionals
under consideration in this article are referred to as {\em
  density-independent} in the literature \cite{wang-carter}. They
  formally read 
\begin{equation}
  \label{eq:nrj}
  T_{\alpha,\beta}[\rho] = \ctf \int_Q \rho^{5/3} + \frac 1 2 \int_Q
  |\nabla \sqrt 
  \rho|^2 + \ctf \int_Q \rho(x)^{\alpha} \, \left( \int_{\RR^3} 
  w_{\alpha,\beta}(k_0[\bar \rho],x-y) \, \rho(y)^{\beta}   \, dy
  \right)  \, dx 
\end{equation}
where $\alpha$ and $\beta$ are positive real numbers such that
  $\alpha+ \beta = 5/3$, where $k_0[ \rho]  = \dps \left( 3 \pi^2
  \bar \rho \right)^{1/3}$ is the Fermi wavenumber associated with the
  average density $\bar \rho = \dps \frac{1}{|Q|} \int_{Q} \rho$ (here
  and below $|Q|$ denotes the volume of the unit cell $Q$), and where
  $w_{\alpha,\beta}$ is some Green 
  kernel. We will denote respectively by $T_{\rm TF}$, $T_{\rm W}$ and
  $T_{\rm C}$ the Thomas-Fermi, von Weizs\"acker and convolution term in
  (\ref{eq:nrj}):
  \begin{equation}
    \label{eq:ektf}
    T_{\rm TF}[\rho] = \ctf \int_Q \rho^{5/3}, 
  \end{equation}
  \begin{equation}
    \label{eq:ekw}
    T_{\rm W}[\rho] = \frac 1 2 \int_Q |\nabla \sqrt \rho|^2
  \end{equation}
  \begin{equation}
    \label{eq:ekc}
    T_{\rm C}[\rho] =  \ctf \int_Q \rho(x)^{\alpha} \, \left( \int_{\RR^3} 
  w_{\alpha,\beta}(k_0[\bar \rho],x-y) \, \rho(y)^{\beta}   \, dy
  \right)  \, dx.
  \end{equation}
Note that $w_{\alpha,\beta}$ is a function of two variables.
  The first one, denoted by $k_F$, is a real number which has the
  dimension of a wavenumber. The second one is the convolution variable; it
  is a vector of $\RR^3$ which has the dimension of a position
  vector. Energy functionals of this type were introduced by Wang 
and Teter \cite{wang-teter} (with $\alpha=\beta= 5/6$), and further
  generalized by several authors \cite{wang-carter, wang-govind-carter,
  wang-govind-carter-2}. For a given pair
  $(\alpha,\beta)$, the Green kernel $w_{\alpha,\beta}$ is completely
  determined by the requirement that the kinetic energy functional
  $T_{\alpha,\beta}$ must be compatible with the Lindhard perturbation
  theory (see e.g. \cite{ashcroft}). This compatibility condition has been
  written as early as in 1964, in the article by Hohenberg and Kohn
  founding the Density-Functional Theory \cite{hohenberg-kohn}. 
Imposing that $T_{\alpha,\beta}$ must be compatible with the
Lindhard theory leads to the relation
$$
\hat w_{\alpha,\beta}(k_F,\xi) = \frac{5}{9\alpha\beta} G\left(
  \frac{|\xi|}{2k_F} \right) 
$$
where $\hat w_{\alpha,\beta}$ denotes the Fourier transform of
$w_{\alpha,\beta}(k_F,x)$ with 
respect to the convolution variable $x$ and where for all $\eta \in \RR^+$,
$$
G(\eta) = \left(\frac12+
  \frac{1-\eta^2}{4\eta}\log\left|\frac{1+
 \eta}{1-\eta} \right| \right)^{-1}  - 3 \eta^2 - 1.
$$
It is important to note that the normalization convention entering in
the definition of the Fourier transform used above is the following:  for
all $f \in L^1(\RR^3)$, 
$$
\hat f(\xi) = \int_{\RR^3} f(x) \, e^{-ix \cdot \xi} \, dx.
$$

\medskip

\noindent
The purpose of this article is to analyze the mathematical properties of
the kinetic energy functionals of the form~(\ref{eq:nrj}). The main
results are presented in Section~\ref{sec:results}. 
We prove that these models are
linearly stable but nonlinearly unstable, in the sense that the
corresponding kinetic energy functionals are not bounded from below. As
a matter of illustration, we provide an example of an electronic density
of simple shape the kinetic energy of which is negative (all the numbers
are in atomic units). Let us consider
a cubical simulation cell $Q = ]-L/2,L/2[^3$ and the $Q$-periodic
function $\rho_{N,r_0,L}$ with $N > 0$ and $r_0 > 0$ defined on $Q$ by
$$
\rho_{N,r_0,L}(x,y,z) = N \, \left( \frac{1}{\pi r_0^2} \right)^{3/2}
e^{-\left(x^2+y^2+z^2\right)/r_0^2}.
$$
For $r_0 << L$, one has (up to machine precision),
$$
\int_Q \rho_{N,r_0,L} = N,
$$
$$
T_{\rm TF} \left[ \rho_{N,r_0,L} \right] = \frac{C_{\rm TF}}\pi \,
\left( \frac 3 5 \right)^{3/2} \, \rho_0^{5/3} \, \frac{L^5}{r_0^2},
$$ 
$$
T_{\rm W} \left[ \rho_{N,r_0,L} \right] = \frac 1 2 \int_Q
|\nabla \sqrt{\rho_{N,r_0,L}}|^2 = \frac 3 2 \, \rho_0 \, \frac{L^3}{r_0^2},
$$
and
$$
T_{\rm C}\left[ \rho_{N,r_0,L} \right] = \frac 5 {9\alpha\beta}  
\, T_{\rm TF} \left[ \rho_{N,r_0,L} \right] \, \phi_{h}(\gamma).
$$
In the above expressions, $T_{\rm TF}$, $T_{\rm W}$ and $T_{\rm C}$ are
defined by (\ref{eq:ektf}), (\ref{eq:ekw}) and (\ref{eq:ekc}), and
\begin{equation} \label{eq:phih}
\phi_{h}(\gamma) = \frac{1}{\pi^{3/2}} \, h^3 \, 
\sum_{q \in (h\ZZ)^3} G(\gamma|q|) \, e^{-|q|^2},
\end{equation}
with 
$$
h = \dps \left( \frac{5}{3 \alpha\beta} \right)^{1/2} \pi
\frac{r_0}L \qquad \mbox{and} \qquad 
\gamma = \dps \left( \frac{3 \alpha\beta}{5} \right)^{1/2} 
 \frac{1}{k_0[\rho_{N,r_0,L}] r_0}.
$$
Notice that (\ref{eq:phih}) is a Riemann sum which approximates 
the integral  
$$
\phi(\gamma) = \frac{1}{\pi^{3/2}} \int_{\RR^3} G(\gamma|q|) \,
e^{-|q|^2} \, dq.
$$
For 
$N=13$, $r_0=0.5$ and $L=4.906$, and with $\alpha,\beta = \dps \frac{5 \pm
  \sqrt{5}}{6}$, one has for instance 
$T_{\alpha\beta}[\rho_{N,r_0,L}] = -8.183$ (for the sake of
comparison, the kinetic energy of the uniform electron gas
of density $\bar\rho=N/L^3$ is $T_{\alpha,\beta}[\bar \rho] = 8.573$). 
The parameters $N=13$ and $L=4.906$ correspond to an all electron
calculation on Aluminium.

\section{Main results} \label{sec:results}


Let us first recall the definitions of the functional spaces under
consideration below: for $1 \le p < +\infty$,
$$L^p_{\rm loc} ({\mathbb R}^3) = \left\{ u \, : \, \RR^3
  \rightarrow \RR \mbox{ measurable}, \quad \int_K |u|^p < +\infty
  \quad \mbox{ for all 
  compact sets $K \subset \RR^3$ } \right\},
$$
$$L^p(Q) =  \left\{ u \, : \, Q
  \rightarrow \RR \mbox{ measurable}, \quad \int_Q |u|^p < +\infty \right\},$$
$$H^1({\mathbb R}^3) = \left\{ u \in L^2(\RR^3), \quad \nabla u \in
  \left( L^2(\RR^3) \right)^3 \right\} ,$$
$$H^1_{\rm per}(Q) = \left\{ u = \left. v \right|_Q, \quad  v \in L_{\rm
    loc}^2(\RR^3) , \quad \nabla v \in  \left(L_{\rm loc}^2(\RR^3)
    \right)^3, \quad v \mbox{ $Q$-periodic}  \right\},$$
$$H^{-1}_{\rm per}(Q) = \left(H^1_{\rm per}(Q)\right)' \mbox{ is the
  topological dual of $H^1_{\rm per}(Q)$}. $$
The space $L^p(Q)$ is equiped with the norm $\| u \|_{L^p} = \dps \left(
  \int_Q |u|^p \right)^{1/p}$ and the space $H^1_{\rm per}(Q)$ with the norm
$\| u \|_{H^1} = \dps  \left(  \int_Q |u|^2 +   \int_Q |\nabla u|^2
  \right)^{1/2}$. 

\medskip

\noindent
The main mathematical properties of the orbital-free kinetic energy
functionals $T_{\alpha,\beta}$ are put together in the following two
theorems. 

\medskip

\begin{theoreme} \label{theo} Let us consider two positive real numbers
  $\alpha$ and $\beta$ such that $\alpha + \beta = 5/3$. Let us consider
  a $Q$-periodic potential $V \in L^{3/2}_{\rm loc}(\RR^3)$ and the
  minimization problem 
\begin{equation} \label{eq:minOF}
I_N = \inf \left\{ T_{\alpha,\beta}[\rho] + \int_{Q} V \rho, \quad 
\rho \ge 0, \quad \sqrt \rho \in H^1_{\rm per}(Q), \quad \int_Q \rho = N
\right\},
\end{equation}
where $N$ is the number of electrons per unit cell. Then,

\begin{enumerate}

\item The real number $T_{\alpha,\beta}[\rho]$ formally defined
  by~(\ref{eq:nrj}) can be rigorously defined for any nonnegative
  function $\rho$ such that $\sqrt{\rho} \in H^1_{\rm per}(Q)$. 

\medskip

\item If $V$ is constant, $\rho_0 = N/|Q|$ is a stable local
  minimizer of~(\ref{eq:minOF}). 

\medskip

\item For $V \in L^{3/2}(Q)$ such that $V - \dps \frac{1}{|Q|} \int_Q V$
  is small enough
  (for the $L^{3/2}(Q)$ norm),
  problem~(\ref{eq:minOF}) has a unique local minimizer in the 
  neighborhood of $\rho_0$. 

\medskip

\item Assume that $V \in L^p(Q)$ with $p>3/2$. When 
$$
N > N_{\alpha,\beta} = \left[ \frac{A_0}{2 \, \ctf
    \left(\frac{8}{9\alpha\beta} - 1 
    \right)} \right]^{3/2},
$$
with 
$$
A_0 = \inf \left\{ \frac{\dps \int_{\RR^3} |\nabla u|^2}
{\dps \int_{\RR^3} |u|^{10/3}} , \quad u \in H^1(\RR^3), \quad 
\int_{\RR^3} u^2 = 1 \right\},
$$
the ground state energy $I_N$ equals $-\infty$. 

\end{enumerate}

\end{theoreme}

\medskip

\begin{figure}[h]
\centering
\psfig{figure=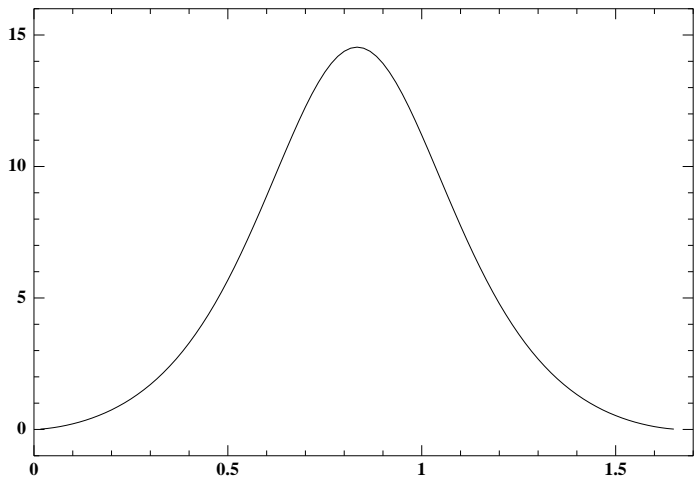,height=7truecm}
\caption{Plot of the function $\alpha \mapsto N_{\alpha,5/3-\alpha}$ for
  $\alpha \in [0,5/3]$.}
\label{fig:Nalphabeta}
\end{figure}

\noindent
It is easy to obtain a numerical value of $A_0$ ($A_0 \simeq 9.5785$),
hence of $N_{\alpha,\beta}$ for all $(\alpha,\beta)$. The 
results are displayed on Fig~\ref{fig:Nalphabeta}. One can see that the
critical 
values $N_{\alpha,\beta}$ are not very large. In particular,
$N_{\alpha,\beta} \simeq 4.636$ for the values recommended in
\cite{wang-carter}, 
namely $\alpha,\beta = \dps \frac{5\pm\sqrt 5}{6}$.

\medskip

\medskip

\noindent
Similar results can be obtained for models in which the electronic
interaction is taken into account, such as the ones used in atomic-scale
simulation of materials. Let us notably consider the
minimization problem
\begin{equation} \label{eq:minOF2}
\widetilde I_N = \inf \left\{ T_{\alpha,\beta}[\rho] + \frac 1 2
  J[\rho-\rho_n] + E_{\rm xc}[\rho]  , \quad  
\rho \ge 0, \quad \sqrt \rho \in H^1_{\rm per}(Q), \quad \int_Q \rho = N
\right\},
\end{equation}
where $\rho_n \in L^{6/5}_{\rm loc}({\mathbb R}^3)$ is a given nonnegative
$Q$-periodic density such that $\dps \int_Q \rho_n = N$ ($\rho_n$
represents the density of smeared nuclear charges), and where $J$ and
$E_{\rm xc}$ are the Coulomb energy functional and some exchange-correlation
energy functional, respectively. Recall that 
$$J[\rho-\rho_n] = \int_Q (\rho-\rho_n) W$$
with $W$ denoting the unique solution in $H^1_{\rm per}(Q)$ of
\begin{equation}
\label{eq:poisson}
\left\{ \begin{array}{l} \dps - \Delta W = 4 \pi (\rho-\rho_n),\\
 \int_Q W = 0. \end{array} \right.
\end{equation}
For simplicity, we consider the case of the so-called X$\alpha$
exchange-correlation functional
$$
E_{\rm xc}[\rho] = - \cxc \int_Q \rho^{4/3},
$$
where $\cxc=\frac34 \left(\frac 3 \pi\right)^{1/3}$ is a positive
constant, but similar results can be 
obtained for more complicated functionals.

\medskip

\begin{theoreme} \label{theo2} Let us consider two positive real numbers
  $\alpha$ and $\beta$ such that $\alpha+ \beta = 5/3$. Then,

\begin{enumerate}

\item If $\rho_n= N/|Q|$ (Jellium background) with $N/|Q| > \rho_{\rm inf}
  \dps \left(\frac{\cxc}{\ctf}\right)^3$, then $\rho_0 =
  N/|Q|$ is a stable local minimizer of~(\ref{eq:minOF2}). 

\medskip

\item For $\rho_n$ $Q$-periodic, such that $\dps \int_Q \rho_n = N$ and
  close enough, for the $L^{6/5}(Q)$ norm, to some constant density
  $\rho_0$ with $\rho_0 > \rho_{\rm inf}$,
  problem~(\ref{eq:minOF2}) has a unique local minimizer in the
  neighborhood of $\rho_0$.

\medskip

\item When
$ \dps
N > N_{\alpha,\beta} = \left[ \frac{A_0}{2 \, \ctf
    \left(\frac{8}{9\alpha\beta} - 1 
    \right)} \right]^{3/2}$,
the ground state energy $\widetilde I_N$ equals $-\infty$. 

\end{enumerate}

\end{theoreme}

\noindent
Note that the constant $\rho_{\rm inf}=\dps
\left(\frac{\cxc}{\ctf}\right)^3 \simeq 0.102$
is not very high ($\bar \rho = 0.110$ for Al and $\bar \rho =
0.309$ for Fe).

\section{Concluding remarks}

\noindent
Density-independent orbital-free kinetic energy functionals of the form
(\ref{eq:nrj}) are all nonlinearly unstable: when the number $N$ of
electrons per unit cell exceeds a few units ($N \ge 5$ for
$\alpha,\beta = \dps \frac{5\pm\sqrt 5}{6}$), the kinetic energy goes to
minus infinity when the density concentrates in some point of the unit
cell. As proved in Theorem~\ref{theo2}, the Coulomb repulsion (which tends to
prevent the density from concentrating) is not able to stabilize the model.
For large, inhomogeneous systems simulated on fine grids, one can
therefore fear that the numerical solution obtained with such models 
will be meaningless. 

\medskip

\noindent
This serious drawback is an additional motivation for constructing more
elaborate functionals such as the so-called {\em density-dependent}
orbital-free functional introduced in \cite{wang-govind-carter-2}. The
mathematical analysis, as well as the numerical simulation of the
latter models, are more difficult. Hopefully, this will be
the matter of a future work.

\section{Mathematical proofs}

\noindent
Let us begin this section by a formal calculation. In the sequel, the
periodic lattice associated to the cell $Q$ is denoted by ${\cal R}$,
and its dual lattice (see e.g. \cite{ashcroft}) by ${\cal R}^\ast$.
If $k_F$ is a positive real number, and if $f$ and $g$ are two
$Q$-periodic functions, one has 
\begin{eqnarray*}
\int_Q f(x) \, \left( \int_{\RR^3}
  w_{\alpha,\beta}(k_F,x-y) \, g(y) \, dy \right) \, dx
& = & \int_Q  f(x) \, \left(  \sum_{k \in {\cal R}} \int_Q    
  w_{\alpha,\beta}(k_F,x-y-k) g(y+k) \, dy \right) \, dx \\
& = & \int_Q  f(x) \, \left(  \sum_{k \in {\cal R}} \int_Q    
  w_{\alpha,\beta}(k_F,x-y-k) g(y) \, dy \right) \, dx \\
& = & \int_Q  \int_Q f(x) \, g(y) \,  \left(  \sum_{k \in {\cal R}}    
  w_{\alpha,\beta}(k_F,x-y-k) \right) \, dx \, dy .
\end{eqnarray*}
As ${\cal R}$ is symmetric with respect to the origin and as
$w_{\alpha,\beta}(k_F,x) = w_{\alpha,\beta}(k_F,-x)$, it follows in
particular that 
$$
\int_Q f(x) \, \left( \int_{\RR^3}
  w_{\alpha,\beta}(k_F,x-y) \, g(y) \, dy \right) \, dx
= \int_Q g(x) \, \left( \int_{\RR^3}
  w_{\alpha,\beta}(k_F,x-y) \, f(y) \, dy \right) \, dx,
$$
and therefore that $T_{\alpha,\beta}[\rho] = T_{\beta,\alpha}[\rho]$. In
addition, using Poisson formula, one obtains
$$
 \sum_{k \in {\cal R}}  w_{\alpha,\beta}(k_F,x-y-k) = \frac{1}{|Q|} \,
\sum_{q \in {\cal R}^\ast}  \hat w_{\alpha,\beta}(k_F,q) e^{i (x-y)
 \cdot q} = \frac{5} {9 \alpha \beta} \, \frac{1}{|Q|} \,
\sum_{q \in {\cal R}^\ast} G \left( \frac{|q|}{2k_F} \right)  
e^{i (x-y) \cdot q} 
$$
then
\begin{eqnarray*}
\int_Q f(x) \, \left( \int_{\RR^3}
  w_{\alpha,\beta}(k_F,x-y) \, g(y) \, dy \right) \, dx
& = & \int_Q  \int_Q f(x) \, g(y) \,  \left( \frac{5} {9 \alpha \beta}
  \, \frac{1}{|Q|} \, 
\sum_{q \in {\cal R}^\ast} G \left( \frac{|q|}{2k_F} \right)  
e^{i (x-y) \cdot q}  \right) \, dx \, dy \\
 & = &  \frac{5} {9 \alpha \beta} \, \frac{1}{|Q|} \,
\sum_{q \in {\cal R}^\ast}  G \left( \frac{|q|}{2k_F} \right)  \,
\overline{c_q(f)} \, c_q(g) 
\end{eqnarray*}
where $(c_q(h))_{q \in {\cal R}^\ast}$ denote the Fourier coefficients
of the $Q$-periodic function $h$, namely
$$
c_q(h) = \int_Q h(x) \, e^{-iq \cdot x} \, dx.
$$ 

\medskip

\noindent{\bf Proof of Theorem~\ref{theo}:} In view of the preceding remark,
we can consider that (\ref{eq:nrj}) is a formal notation for
\begin{equation} \label{eq:Tbon}
T_{\alpha,\beta}[\rho] =   \ctf \int_Q \rho^{5/3} + \frac 1 2 \int_Q
  |\nabla \sqrt \rho|^2 + \ctf 
\frac{5} {9 \alpha \beta} \, \frac{1}{|Q|} \,
\sum_{q \in {\cal R}^\ast}  G \left( \frac{|q|}{2k_0[\rho]} \right)  \,
\overline{c_q\left(\rho^\alpha\right)} \, c_q\left(\rho^\beta\right). 
\end{equation}
Note that the latter expression of the nonlocal term is the one which is
actually used in numerical simulations (see
e.g. \cite{wang-govind-carter-2}). Now, it is easy to see that
$T_{\alpha,\beta}[\rho]$ is well defined by (\ref{eq:Tbon}) for any
nonnegative function $\rho$ such that $\sqrt \rho \in H^1_{\rm per}(Q)$ as
soon as $\alpha$ and $\beta$ are positive real numbers such that $\alpha
+ \beta = \dps \frac 5 3$. Indeed, when $\sqrt \rho \in H^1_{\rm per}(Q)$, both
$\rho^{5/3}$ and $|\sqrt \rho|^2$ are in $L^1(Q)$. Besides,
the sum over the dual lattice is normally convergent. This
can be proved by remarking that, on the one hand, $G$
is a bounded function, and that, on the other hand, $\rho^\alpha$ and
$\rho^\beta$ are in $L^2(Q)$ for $\sqrt{\rho} \in H^1_{\rm per}(Q)$ and
for $\alpha$ and $\beta$ are in $[0,5/3]$ (these are consequences of Sobolev
inequalities~\cite{zeidler}). Thus, using
Cauchy-Schwarz inequality and Parseval relation, 
\begin{eqnarray*}
\frac{1}{|Q|} \,
\sum_{q \in {\cal R}^\ast}  \left| G \left( \frac{|q|}{2k_0[\rho]} \right)  \,
\overline{c_q\left(\rho^\alpha\right)} \, c_q\left(\rho^\beta\right)
\right| & \le & \left( \sup_{\RR^+} |G| \right) \,
\frac{1}{|Q|} \sum_{q \in {\cal R}^\ast} \left|
  \overline{c_q\left(\rho^\alpha\right)} \, c_q\left(\rho^\beta\right) 
\right|
\\ & \le & \left( \sup_{\RR^+} |G| \right) \,
\frac{1}{|Q|} \left( \sum_{q \in {\cal R}^\ast} \left|
  \overline{c_q\left(\rho^\alpha\right)} \right|^2 \right)^{1/2}
\,   \left( \sum_{q \in {\cal R}^\ast} \left| c_q\left(\rho^\beta\right)
  \right|^2 \right)^{1/2}
\\
& = &
\left( \sup_{\RR^+} |G| \right) \, \| \rho^{\alpha} \|_{L^2(Q)}
\,  \| \rho^{\beta} \|_{L^2(Q)} < + \infty.
\end{eqnarray*}
This concludes the proof of the first statement of Theorem~\ref{theo}.
Let us now prove the second statement. For this
purpose we introduce the problem 
\begin{equation} \label{eq:phi}
\inf \left\{ E_K[\phi] + \int_{Q} V \phi^2, \quad 
\quad \phi \in H^1_{\rm per}(Q), \quad \int_Q \phi^2 = N
\right\},
\end{equation}
where 
$$
E_K[\phi] = T_{\alpha,\beta}[\phi^2] =  
\ctf \int_Q |\phi|^{10/3} + \frac 1 2 \int_Q
  |\nabla \phi|^2 + \ctf 
\frac{5} {9 \alpha \beta} \, \frac{1}{|Q|} \,
\sum_{q \in {\cal R}^\ast}  G \left( \frac{|q|}{2k_0[\rho_0]} \right)  \,
\overline{c_q\left(|\phi|^{2\alpha}\right)} \,
c_q\left(|\phi|^{2\beta}\right).  
$$
It is easy to check that the infimum of~(\ref{eq:phi}) is equal to $I_N$ and
that $\rho$ is a local minimum of~(\ref{eq:minOF})
if and only if $\phi= \sqrt{\rho}$ is a local minimum
of~(\ref{eq:phi}). The Euler-Lagrange equation associated with problem
(\ref{eq:phi}) reads
\begin{equation} \label{eq:ELphi}
- \frac 1 2  \Delta \phi + \frac 5 3 \ctf
|\phi|^{4/3} \phi   +  V \phi 
+ L[\phi]  = \mu  \phi 
\end{equation}
where $L[\phi]$ denotes the continuous linear form on $H^1_{\rm per}(Q)$
defined by 
\begin{equation}
\label{eq:lin}
L[\phi] \cdot h = \ctf \frac{5} {9 \alpha \beta} \, \frac{1}{|Q|} \,
\sum_{q \in {\cal R}^\ast}  G \left( \frac{|q|}{2k_0[\rho_0]} \right)  \,
\left( \alpha \overline{c_q\left(|\phi|^{2\alpha-2} \phi h \right)} \,
c_q\left(|\phi|^{2\beta}\right) 
+ \beta \overline{c_q\left(|\phi|^{2\alpha}\right)} \,
c_q\left(|\phi|^{2\beta-1} \phi h \right) \right),
\end{equation}
and where $\mu$ is a Lagrange multiplier associated with the
constraint $\dps \int_Q \phi^2 = N$. Let us denote by $\phi_0 =
\sqrt{\rho_0}$. As on the one hand, 
$c_q\left(\phi_0^{2\alpha}\right) =
c_q\left(\phi_0^{2\beta}\right) =0$ for all $q \neq 0$, and as on the
other hand, $ G \left(0 \right) = 0$, one has $L[\phi_0] =
0$. Therefore, $\phi_0$ solves the Euler-Lagrange equation
(\ref{eq:ELphi}) for $V$ equal to the constant $V_0$, with $\mu = \frac 5 3
\rho_0^{2/3} + V_0$; if $V$ is constant,
$\phi_0$ is thus a critical point of (\ref{eq:phi}). In order to
complete the proof of the second statement of Theorem~\ref{theo}, it is
sufficient to show that the continuous symmetric bilinear form 
\begin{equation} \label{eq:defB}
B[\phi_0,\mu_0](h_1,h_2) = \frac 1 2 \int_Q \nabla h_1 \cdot \nabla h_2
+ \frac{35}9 \ctf \int_Q \phi_0^{4/3} h_1h_2 + \frac{5}{9\alpha\beta}
\ctf {\cal K}[\phi_0](h_1,h_2) - \mu_0 \int_Q h_1 h_2 ,
\end{equation}
is positive definite on the tangent subspace $\left\{ h \in
  H^1_{\rm per}(Q), \; \dps \int_Q \phi_0 h = 0 \right\}$. In the above
expression $\mu_0 = \frac 5 3 \rho_0^{2/3}$ and ${\cal K}[\phi_0]$
  denotes the bilinear form defined by 
\begin{eqnarray*}
{\cal K}[\phi_0](h_1,h_2) & = & \frac{1}{|Q|} \,
\sum_{q \in {\cal R}^\ast}  G \left( \frac{|q|}{2k_0[\rho_0]} \right)  \,
\left( \alpha (2\alpha-1) \,  \overline{c_q\left(\phi_0^{2\alpha-2} h_1h_2
    \right)} \, 
c_q\left(\phi_0^{2\beta}\right) 
+ \beta (2\beta-1) \,  \overline{c_q\left(\phi_0^{2\alpha}\right)} \,
c_q\left(\phi_0^{2\beta-2} h_1 h_2 \right) \right.
\\ & & + \left. 2 \alpha \beta  \left( \overline{c_q\left(\phi_0^{2\alpha-1}
      h_1\right)} c_q\left(\phi_0^{2\beta-1} h_2 \right)  
+  \overline{c_q\left(\phi_0^{2\alpha-1}
      h_2\right)} c_q\left(\phi_0^{2\beta-1} h_1 \right)  \right) \right) \\
& = & \frac{4 \alpha \beta}{|Q|} \, \phi_0^{2\alpha+ 2 \beta -2} 
\sum_{q \in {\cal R}^\ast} G \left( \frac{|q|}{2k_0[\rho_0]} \right)
\, \mbox{Re } \left( \overline{c_q\left( h_1\right)} c_q\left(h_2
  \right) \right) \\
&=&\frac{4 \alpha \beta}{|Q|} \, \phi_0^{4/3}\, 
\sum_{q \in {\cal R}^\ast} G \left( \frac{|q|}{2k_0[\rho_0]} \right)
\, \mbox{Re } \left( \overline{c_q\left( h_1\right)} c_q\left(h_2
  \right) \right) .
\end{eqnarray*}
The latter equality has been obtained using that for all $q \neq 0$, 
$c_q\left(\phi_0^{2\alpha}\right) =
c_q\left(\phi_0^{2\beta}\right)=0$, that $G(0) = 0$, and that
$\alpha+\beta = 5/3$. A simple calculation then leads to
\begin{eqnarray*}
B[\phi_0,\mu_0](h,h) & = & \frac 1 2 \int_Q |\nabla h|^2 + \frac{20}{9|Q|} \ctf
\phi_0^{4/3} \sum_{q \in {\cal R}^\ast} \left[ G \left(
  \frac{|q|}{2k_0[\rho_0]} \right) + 1 \right] 
\overline{c_q\left(h\right)} c_q\left(h \right) \\
& = & \frac 1 2 \int_Q |\nabla h|^2 + \frac 2 {3|Q|} \, \left( k_0[\rho_0]
\right)^2 \, \sum_{q \in {\cal R}^\ast} \left[ G \left(
  \frac{|q|}{2k_0[\rho_0]} \right) + 1 \right] 
\overline{c_q\left(h\right)} c_q\left(h \right) \\
& = & \frac 2 {3|Q|} \, \left( k_0[\rho_0]
\right)^2 \, \sum_{q \in {\cal R}^\ast} \left[ G \left(
  \frac{|q|}{2k_0[\rho_0]} \right) + 1 - 3 \left(
  \frac{|q|}{2k_0[\rho_0]} \right)^2 \right] 
\overline{c_q\left(h\right)} c_q\left(h \right) \\
& = & \frac 2 {3|Q|} \, \left( k_0[\rho_0]
\right)^2 \, \sum_{q \in {\cal R}^\ast} F \left(
  \frac{|q|}{2k_0[\rho_0]} \right)
\overline{c_q\left(h\right)} c_q\left(h \right) 
\end{eqnarray*}
where the function $F$ is defined by
$$
F(\eta) =  \left(\frac 1 2 + \frac{1-\eta^2}{4\eta}\log\left|\frac{1+
 \eta}{1-\eta} \right| \right)^{-1}.
$$
Not surprisingly, one recovers the function $F(\eta)$ arising in Lindhard
theory~\cite{ashcroft}. As $F(\eta) \ge 1$ for all $\eta \ge 0$, one
obtains 
\begin{equation} \label{eq:coercive}
B[\phi_0,\mu_0](h,h) \ge \frac 2 {3} \, \left( k_0[\rho_0]\right)^2
\int_Q h^2,
\end{equation}
which completes the proof of the second statement. 

\medskip

\noindent
Let us now consider the function
$$
\begin{array}{rccl}
{\cal F}\; : & \dps \left( H^1_{\rm per}(Q) \times \RR \right) \times L^{3/2}(Q)
& \longrightarrow & \left( H^{-1}_{\rm per}(Q) \times \RR \right) \\
& \dps \left( \left(\phi,\mu \right), V \right) & \longmapsto &
\dps \left( 
-\frac 1 2 \Delta \phi +  \frac 5 3 |\phi|^{4/3} \phi  +  V \phi 
+ L[\phi] - \mu \phi , \int_{Q} \phi^2 - N \right) .
\end{array}
$$
The function ${\cal F}$ is of class $C^1$ and satisfies, for any constant $V_0$,
$\dps {\cal F} \left( \left(\phi_0,\mu_0+V_0  \right), V_0 \right) = 0$. Besides,
the partial derivative of $F$ with respect to the pair $(\phi,\mu)$, at
the point $\left( \left(\phi_0,\mu_0+V_0 \right), V_0 \right)$ is given by
$$
\left. \frac{\partial {\cal F}}{\partial \phi} \right|_{\left(
    \left(\phi_0,\mu_0 + V_0 \right), V_0 \right)} \cdot (h) 
= \left( B[\phi_0,\mu_0] \left( h,\cdot \right), \int_{Q} \phi_0 h
\right), 
$$
$$\left.\frac{\partial {\cal F}}{\partial \mu} \right|_{\left(
    \left(\phi_0,\mu_0 + V_0 \right), V_0 \right)} = (-\phi_0,0),$$
where $B[\phi_0,\mu_0]$ denotes the bilinear form defined by
(\ref{eq:defB}). Next, it is possible to improve (\ref{eq:coercive}), by
showing that there exists some constant $\gamma>0$ such that $F(\eta)
\geq 1+\gamma \eta^2$ for all $\eta\geq 0$ (actually, one can use
$\gamma = 1/4$ ). Hence, we have
$$B[\phi_0, \mu_0](h,h) \geq \frac 2 3 (k_0[\rho_0])^2 \int_Q h^2
  + \frac \gamma 6 \int_Q |\nabla h|^2.$$
This shows that $B[\phi_0, \mu_0]$ is coercive, and we may thus apply
Lax-Milgram theorem \cite{FA}, proving that  $\dps
  \left. \frac{\partial {\cal F}}{\partial (\phi,\mu)} \right|_{\left(
  \left(\phi_0,\mu_0 +V_0\right), V_0 \right)}$ is invertible. The third
  statement of Theorem~\ref{theo} then follows from the implicit
  function theorem~\cite{FA}. 

\medskip

\noindent
The fourth statement can be established by a scaling argument. We choose
the coordinate axes in such a way that $B(0,\epsilon) = \left\{ x \in
  \RR^3, \; |x| < \epsilon \right\} \subset Q$ for
some $\epsilon > 0$, and we consider
a density $\rho_1 \in C^\infty_0(\RR^3)$ supported in $B(0,\epsilon)$
such that $\rho_1 \ge 0$, $\dps \int_{\RR^3} \rho_1 = N$. We then
consider the family of trial densities $(\rho_\sigma)_{\sigma \ge 1}$
defined by 
$$
\forall \sigma \ge 1, \quad \forall x \in Q, \quad  \rho_\sigma(x) = 
\sigma^3 \, \rho_1(\sigma x).
$$
It is clear that for all $\sigma \ge 1$, $\rho_\sigma$ belongs to the
minimization set
$$
\left\{ \rho \ge 0, \quad \sqrt{\rho} \in H^1_{\rm per}(Q), \quad
\int_Q \rho = N \right\}.
$$
One has
$$
T_{\alpha,\beta}[\rho_\sigma] =   \ctf \int_Q \rho_\sigma^{5/3} + \frac
1 2 \int_Q |\nabla \sqrt \rho_\sigma|^2 + \ctf 
\frac{5} {9 \alpha \beta} \, \frac{1}{|Q|} \,
\sum_{q \in {\cal R}^\ast}  G \left( \frac{|q|}{2k_0[\rho]} \right)  \,
\overline{c_q\left(\rho_\sigma^\alpha \right)} \,
c_q\left(\rho_\sigma^\beta\right).  
$$
As
$$ 
\int_Q \rho_\sigma^{5/3} = \sigma^2  \int_{\RR^3} \rho_1^{5/3},
\qquad
\frac 1 2 \int_Q |\nabla \sqrt \rho_\sigma|^2 = 
\frac {\sigma^2} 2 \int_{\RR^3} |\nabla \sqrt \rho_1|^2, 
$$
$$
c_q\left(\rho_\sigma^\alpha \right) = \sigma^{3(\alpha-1)} \;
\widehat{\rho_1^\alpha} \left( \frac{q}{\sigma} \right),  \quad
\mbox{and} \quad
c_q\left(\rho_\sigma^\beta \right) =  \sigma^{3(\beta-1)}  \;
\widehat{\rho_1^\beta} \left( \frac{q}{\sigma} \right),
$$
we obtain
$$
T_{\alpha,\beta}[\rho_\sigma] =  \sigma^2 \,
\left[ \frac 1 2 \int_{\RR^3} |\nabla \sqrt \rho_1|^2
+ \ctf \int_{\RR^3} \rho_1^{5/3}
+  \ctf 
\frac{5} {9 \alpha \beta} \, \frac{1}{\sigma^3 |Q|} \,
\sum_{q \in {\cal R}^\ast}  G \left( \frac{|q|}{2k_0[\rho]} \right)  \,
\overline{\widehat{\rho_1^\alpha} \left( \frac{q}{\sigma} \right)} \,
\widehat{\rho_1^\beta} \left( \frac{q}{\sigma} \right)
\right].
$$
Using that $G$ is bounded and that $\dps \lim_{\eta \rightarrow +\infty}
G(\eta) = - \frac 8 5$, we have  
\begin{eqnarray*}
\frac{1}{\sigma^3 |Q|} \,
\sum_{q \in {\cal R}^\ast}  G \left( \frac{|q|}{2k_0[\rho]} \right)  \,
\overline{\widehat{\rho_1^\alpha} \left( \frac{q}{\sigma} \right)} \,
\widehat{\rho_1^\beta} \left( \frac{q}{\sigma} \right) & = &
\frac{1}{\sigma^3 |Q|} \,
\sum_{q \in \frac 1 \sigma {\cal R}^\ast}  G \left( \frac{\sigma
    |q|}{2k_0[\rho]} \right)  \, 
\overline{\widehat{\rho_1^\alpha} \left( q \right)} \,
\widehat{\rho_1^\beta} \left( q \right) \\
& \dps \mathop{\longrightarrow}_{\sigma \rightarrow +\infty} &
- \frac{8}{5} \frac{1}{(2\pi)^3} 
\int_{\RR^3} \overline{\widehat{\rho_1^\alpha} \left( q \right)} \, 
\widehat{\rho_1^\beta} \left( q \right) \, dq \\
& = & - \frac{8}{5} \int_{\RR^3} \rho_1^{5/3}.
\end{eqnarray*}
As $N > N_{\alpha,\beta}$, and as $C^\infty_0(\RR^3)$ is dense in
$H^1(\RR^3)$, it is possible to find a function $\phi \in
C^\infty_0(\RR^3)$ such that $\dps \int_{\RR^3} \phi^2 = 1$ and
\begin{equation}\label{eq:bphi}
\frac{\dps \int_{\RR^3} |\nabla \phi|^2}{\dps \int_{\RR^3} |\phi|^{10/3}}
< 2 \ctf \left( \frac{8}{9\alpha\beta} - 1 \right) \, N^{2/3}.
\end{equation}
For some $\tau$ large enough, the function 
$$
\phi_\tau(x) = \tau^{3/2} \phi(\tau x)
$$
is supported in the set $B(0,\epsilon)$ introduced above and the
function $\rho_1(x) = N \phi_\tau(x)^2$ is such that 
 $\rho_1 \in C^\infty_0(\RR^3)$,  ${\rm Supp} \, \rho_1 \subset
 B(0,\epsilon)$, $\rho_1 \ge 0$ and $\dps \int_{\RR^3} \rho_1 = N$.
In addition, 
$$
\gamma = -  \frac 1 2 \int_{\RR^3} |\nabla \sqrt{\rho_1}|^2
+ \ctf \left( \frac{8}{9\alpha\beta} - 1 \right) \int_{\RR^3}
|\rho_1|^{5/3} > 0.
$$
One therefore has,
$$
T_{\alpha,\beta}[\rho_\sigma] \mathop{\sim}_{\sigma
  \rightarrow +\infty}  - \gamma \sigma^2.
$$
Besides, from H\"older inequality \cite{FA},
$$
\left| \int_Q  \rho_{\sigma} V  \right| \le  
\| \rho_\sigma \|_{L^{p'}} \, \| V \|_{L^p}
$$
where $p' = \dps \left( 1 - \frac 1 p \right)^{-1} < 3$.
As $\| \rho_\sigma \|_{L^{p'}} = \sigma^{3-3/p'} \, \| \rho_1
\|_{L^{p'}} = o(\sigma^2)$, we finally conclude that 
$$
T_{\alpha,\beta}[\rho_\sigma] + \int_Q \rho_\sigma V  \mathop{\sim}_{\sigma
  \rightarrow +\infty}  - \gamma \sigma^2,
$$ 
and therefore that $I_N = - \infty$.

\bigskip

\begin{remarque}
 Let us point out that one can carry out the same analysis with the density
\begin{equation} \label{eq:rhosigma}
\rho_\sigma(x) = \frac{N-N_c}{|Q|} + N_c \sigma^3
\rho_1(\sigma x),
\end{equation}
with $N_{\alpha,\beta}<N_c<N$ instead of the above $\rho_\sigma$. This is
physically more satisfactory since the densities defined by
(\ref{eq:rhosigma}) are uniformly bounded away from zero. 
\end{remarque}

\bigskip

\noindent{\bf Proof of Theorem~\ref{theo2}:} We use the same strategy as
in the proof of Theorem~\ref{theo}, and thus define
the following minimization problem:
\begin{equation} \label{eq:phi2}
\inf \left\{ E_K[\phi] + \frac12 J[\phi^2 - \rho_n] + E_{\rm xc}[\phi^2], \quad 
\quad \phi \in H^1_{\rm per}(Q), \quad \int_Q \phi^2 = N
\right\}.
\end{equation}
A function $\phi$ is a solution of (\ref{eq:phi2}) if and only if $\rho
= \phi^2$ is a solution of (\ref{eq:minOF2}). Let us write down the
corresponding Euler-Lagrange equation:
\begin{equation}
  \label{eq:ELphi2}
  -\frac 1 2  \Delta \phi + \frac 5 3\ctf
|\phi|^{4/3} \phi + L[\phi] + W\phi -\frac43 \cxc |\phi|^{2/3}\phi =
\mu  \phi , 
\end{equation}
where the electrostatic potential $W$ is defined by (\ref{eq:poisson}),
and the linear form $L[\phi]$ by (\ref{eq:lin}). The constant $\mu$ is
the Lagrange multiplier associated to the charge constraint. As pointed
out in the proof of Theorem~\ref{theo}, if we define $\phi_0 =
\sqrt{\rho_0}$, we have $L[\phi_0]=0$. In addition, $W$ is then a
solution of $-\Delta W = 0$, and is thus identically zero in view of its
periodicity and of the normalization condition in
(\ref{eq:poisson}). This shows that $\phi_0$ is a solution of
(\ref{eq:ELphi2}) with 
\begin{equation}
\label{eq:mult}
\mu = \mu_0 = \frac53 \ctf \rho_0^{2/3} -
\frac43 \cxc \rho_0^{1/3}.
\end{equation}
Hence, $\phi_0$ is a criticial point of the energy. We need now to show
that it is a local minimizer. In order to do so, we show that the
bilinear form
\begin{eqnarray} \label{eq:defB2}
B[\phi_0,\mu_0](h_1,h_2) &=& \frac 1 2 \int_Q \nabla h_1 \cdot \nabla h_2
+ \frac{35}9 \ctf \int_Q \phi_0^{4/3} h_1h_2 + \frac{5}{9\alpha\beta}
\ctf {\cal K}[\phi_0](h_1,h_2) \nonumber \\
&&+ 2\phi_0 D(h_1,h_2)- \cxc\frac{20}{9} \int_Q \phi_0^{2/3} h_1 h_2- \mu_0 \int_Q h_1 h_2 ,
\end{eqnarray}
is positive definite on the tangent subspace $\left\{ h \in
  H^1_{\rm per}(Q), \; \dps \int_Q \phi_0 h = 0 \right\}$. In the above
expression $\mu_0$ is defined by (\ref{eq:mult}), ${\cal K}[\phi_0]$
  denotes the bilinear form defined by 
$$
{\cal K}[\phi_0](h_1,h_2) =\frac{4 \alpha \beta}{|Q|} \, \phi_0^{4/3}\, 
\sum_{q \in {\cal R}^\ast} G \left( \frac{|q|}{2k_0[\rho_0]} \right)
\, \mbox{Re } \left( \overline{c_q\left( h_1\right)} c_q\left(h_2
  \right) \right),
$$
and $D$ is defined by
$$D(h_1,h_2) = \int_Q W_1 h_2, \mbox{ with } -\Delta W_1 = 4\pi h_1, \quad
W_1 \in H^1_{\rm per}(Q), \quad \int_Q W_1 = 0.$$
We now point out that actually, $W_1$ may be defined by its Fourier
coefficients through $|q|^2c_q(W_1) = 4\pi c_q(h_1)$ and $c_0(W_1)=0$, so that
$$D(h_1,h_2) = \frac{4\pi}{|Q|} \sum_{q\in {\cal R}^*\setminus\{0\}}
\frac{c_q(h_1) \overline{c_q(h_2)}}{|q|^2}.$$
Hence, carrying out the same computation as in the proof of
Theorem~\ref{theo}, we have
\begin{eqnarray*}
B[\phi_0,\mu_0](h,h) &=& \frac 1 {|Q|} \,
 \sum_{q \in {\cal R}^\ast} \left(\frac{20}9\ctf \rho_0^{2/3} F \left(
  \frac{|q|}{2k_0[\rho_0]} \right)-\frac{20}9 \cxc \rho_0^{1/3} \right)
 |c_q(h)|^2 \\
&& + \frac{1}{|Q|} \sum_{q\in {\cal R}^*\setminus\{0\}}
 \frac{4\pi}{|q|^2} |c_q(h)|^2.
\end{eqnarray*}
Now, we know that $F(\eta)\geq 1$, which, with the help of $\rho_0^{1/3}
> \dps \frac \cxc \ctf$, implies that 
$$B[\phi_0, \mu_0](h,h) \geq \delta \int_Q h^2,$$
for some positive constant $\delta$ independent of $h$. This proves that
$\rho_0$ is a local minimizer of (\ref{eq:minOF2}).

We now prove the second statement of Theorem~\ref{theo2}. For this
purpose, we introduce the function
$$
\begin{array}{rccl}
{\cal G}\; : & \dps \left( H^1_{\rm per}(Q) \times \RR \right) \times L^{6/5}(Q)
& \longrightarrow & \left( H^{-1}_{\rm per}(Q) \times \RR \right) \\
& \dps \left( \left(\phi,\mu \right), \rho_n \right) & \longmapsto &
\dps \left( 
-\frac 1 2 \Delta \phi +  \frac 5 3 \ctf|\phi|^{4/3} \phi  
+ L[\phi] +  W \phi  - \frac43 \cxc |\phi|^{2/3}\phi- \mu \phi ,
\int_{Q} \phi^2 - N \right), 
\end{array}
$$
where $W$ is here again defined by (\ref{eq:poisson}), where $\rho = |\phi|^2.$
The function ${\cal G}$ is of class $C^1$ (all terms are clearly $C^1$
except $W\phi$, but this one may be shown to have the desired
regularity with the help of standard elliptic estimates). In addition,
${\cal G}((\phi_0,\mu_0),\rho_0) = 0,$ and the partial derivatives of
${\cal G}$ at this point read
$$
\left. \frac{\partial {\cal G}}{\partial \phi} \right|_{\left(
    \left(\phi_0,\mu_0  \right), \rho_0 \right)} \cdot (h) 
= \left( B[\phi_0,\mu_0] \left( h,\cdot \right), \int_{Q} \phi_0 h
\right), 
$$
$$\left.\frac{\partial {\cal G}}{\partial \mu} \right|_{\left(
    \left(\phi_0,\mu_0  \right), \rho_0 \right)} = (-\phi_0,0),$$
where $B[\phi_0,\mu_0]$ is defined by (\ref{eq:defB2}). Here again,
    using the fact that $F(\eta) \geq 1+\gamma \eta^2$ for some positive
    constant $\gamma$ and that $\rho_0^{1/3} > \frac\cxc\ctf,$ one
    easily shows that 
$$B[\phi_0,\mu_0](h,h) \geq \delta \left(\int_Q h^2 + \int_Q |\nabla
  h|^2\right),$$
for some constant $\delta>0$. Hence, one may apply Lax-Milgram theorem
  \cite{FA} to prove that $\dps
  \left. \frac{\partial {\cal G}}{\partial (\phi,\mu)} \right|_{\left(
  (\phi_0,\mu_0 ), \rho_0 \right)}$ is invertible. The second
  statement of Theorem~\ref{theo} then follows from the implicit
  function theorem~\cite{FA}. 

\medskip

\noindent
We now turn to the third statement of Theorem~\ref{theo2}. We use
test functions of the form
$$\rho_\sigma(x) = \sigma^3\rho_1(\sigma x), \quad \sigma\geq 1.$$
We can carry out the same computation for the kinetic energy, showing
here again that, if $\phi_1$ satisfies (\ref{eq:bphi}), choosing
$\rho_1(x) = N |\phi_1(x)|^2$ leads to 
$$
T_{\alpha,\beta}[\rho_\sigma] \mathop{\sim}_{\sigma
  \rightarrow +\infty}  - \gamma \sigma^2,
$$
with
$$
\gamma = -  \frac 1 2 \int_{\RR^3} |\nabla \sqrt{\rho_1}|^2
+ \ctf \left( \frac{8}{9\alpha\beta} - 1 \right) \int_{\RR^3}
|\rho_1|^{5/3} > 0.
$$
We therefore only need to check that the remaining terms of the energy
have a scaling of lower order as $\sigma$ goes to infinity. First, we
have
$$\int_Q \rho_\sigma^{4/3} = \int_Q \sigma^4 \rho_1^{4/3}(\sigma x) dx =
\sigma \int_{{\mathbb R}^3} \rho_1^{4/3}.$$
We then compute the Coulomb term, using its Fourier expression:
\begin{eqnarray*}
J(\rho_\sigma - \rho_n) &=& \frac{4\pi}{|Q|} \sum_{q\in {\cal
    R}^*\setminus\{0\}} \frac{|c_q(\rho_\sigma - \rho_n)|^2}{|q|^2}
\leq \frac{8\pi}{|Q|} \sum_{q\in {\cal
    R}^*\setminus\{0\}} \frac{|c_q(\rho_\sigma)|^2 +
    |c_q(\rho_n)|^2}{|q|^2}.
\end{eqnarray*}
Pointing out, as in the proof of Theorem~\ref{theo}, that
$c_q(\rho_\sigma) = \hat\rho_1\left(\frac{q}{\sigma}\right),$ we thus
have
$$J(\rho_\sigma - \rho_n) \leq \frac{8\pi}{|Q|\sigma^2} \sum_{q\in{\cal
    R}^*\setminus\{0\}}
    \frac{\left|\hat\rho\left(\frac{q}{\sigma}\right)\right|^2}{\frac{|q|^2}{\sigma^2}}
    + C,$$
where $C$ is a constant depending only on $\rho_n$. The sum is, up to a
    factor $\sigma$, a Riemann sum, and we thus have
$$J(\rho_\sigma - \rho_n) \leq 2\sigma \int_{{\mathbb R}^3} \frac{|\hat
  \rho_1(\xi)|^2}{|\xi|^2}d\xi + o(\sigma).$$
This allows to conclude that both the exchange term and the
  electrostatic term have a scaling of order strictly lower than
  $\sigma^2$ as $\sigma$ goes to infinity. We thus come to the same
  conclusion as in Theorem~\ref{theo}.

\nocite{*}
\bibliographystyle{plain}
\bibliography{articleOF}

\end{document}